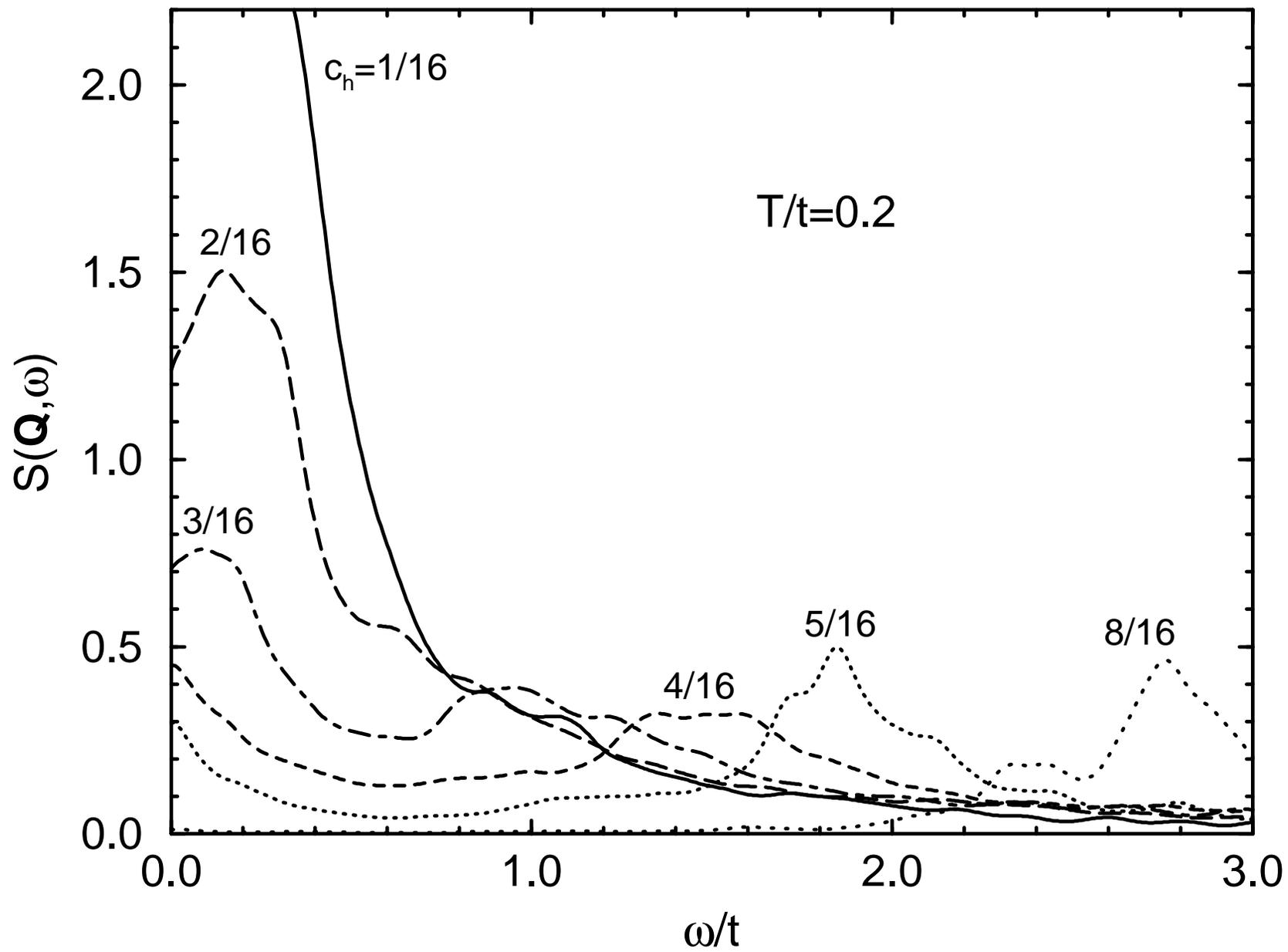

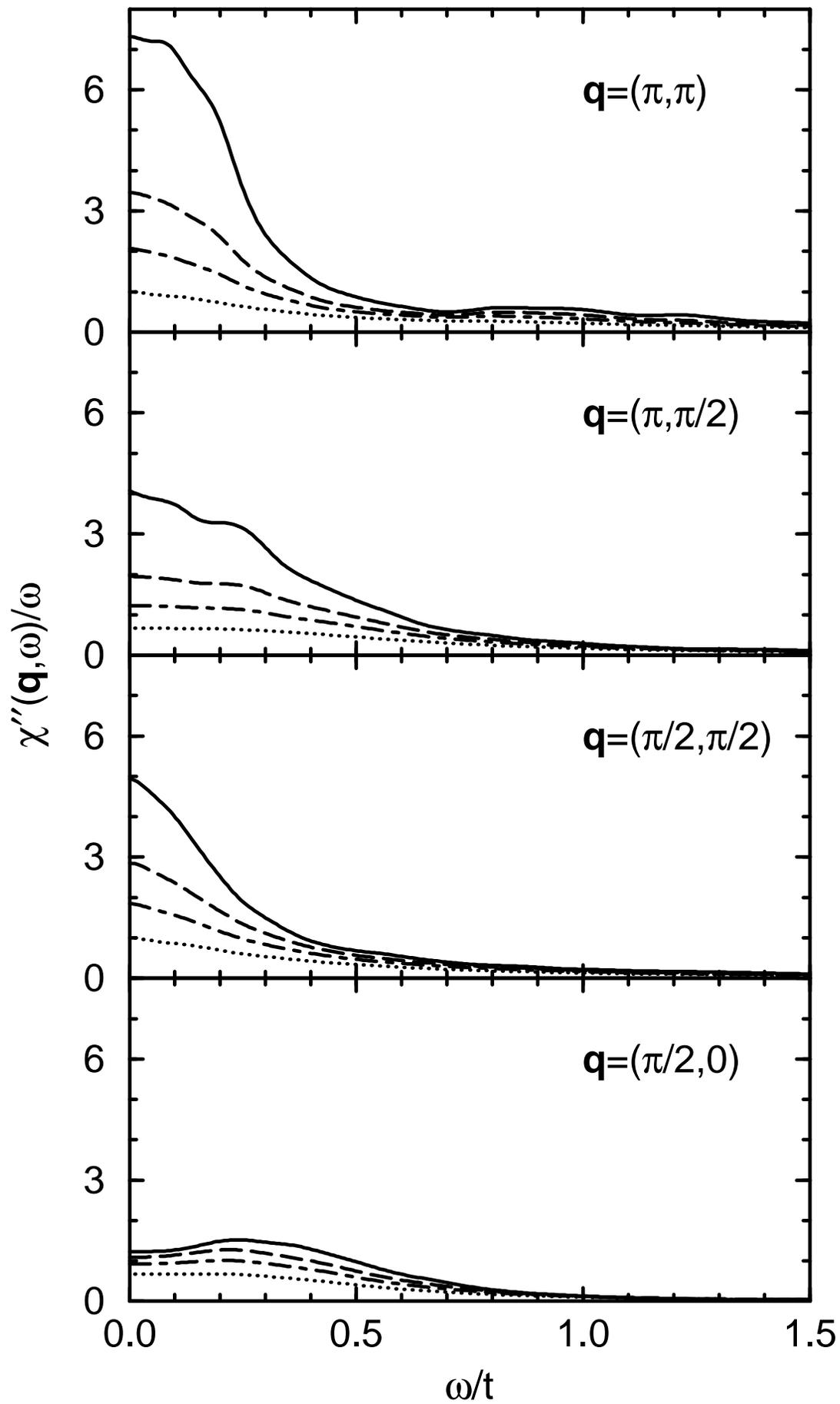

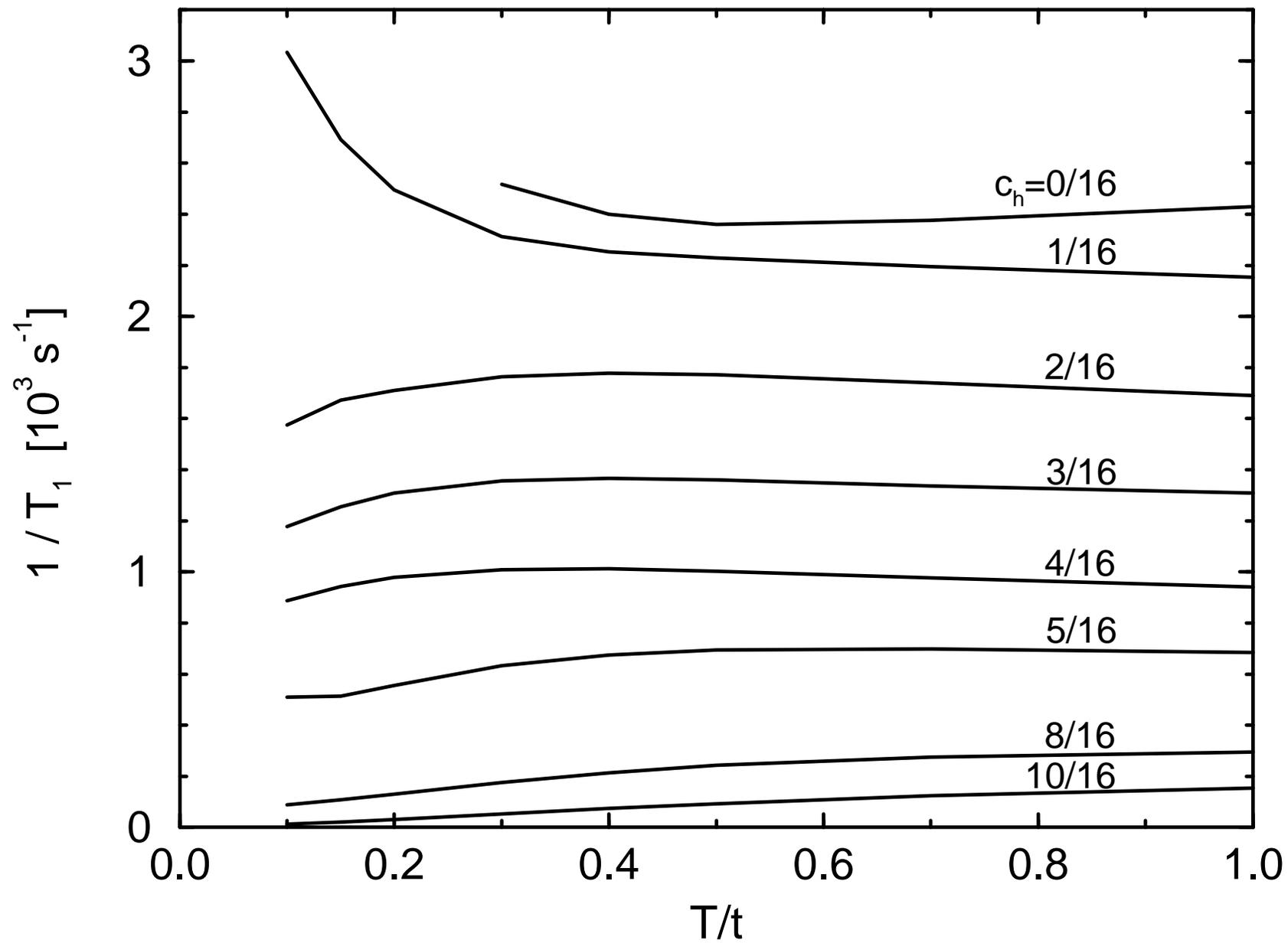

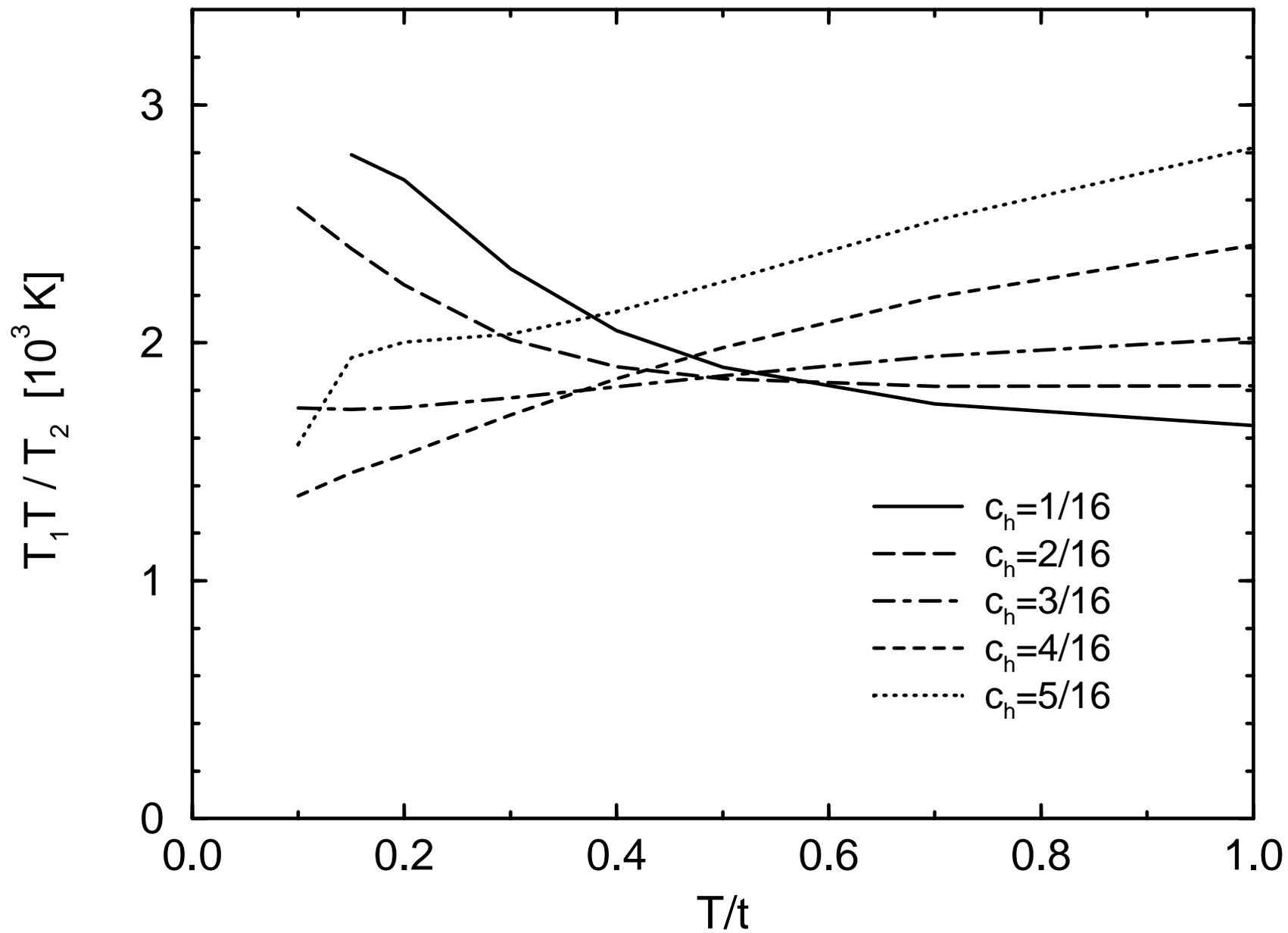

# Anomalous Spin Dynamics in Doped Quantum Antiferromagnets

J. Jaklič and P. Prelovšek

*J. Stefan Institute, University of Ljubljana, 61111 Ljubljana, Slovenia*

Finite-temperature spin dynamics in planar $t - J$ model is studied using the method based on the Lanczos diagonalization of small systems. Dynamical spin structure factor at moderate dopings shows the coexistence of free-fermion-like and spin-fluctuation timescales. At $T < J$, the low-frequency and static susceptibility show pronounced $T$ dependence, supporting a scenario, related to the marginal Fermi-liquid one, for the explanation of neutron-scattering and NMR-relaxation experiments in cuprates. Calculated NMR relaxation rates reasonably reproduce experimental ones.

The understanding of spin and charge dynamics in strongly correlated systems, as realized in cuprates on doping the reference antiferromagnetic (AFM) insulator, still represents a challenge for theoreticians. The low-frequency spin dynamics and static spin response in the undoped and doped AFM state in cuprates have been recently extensively studied by neutron scattering [1] and in a number of NMR and NQR experiments [2,3]. They established that normal-state spin dynamics differs qualitatively from the one expected for Landau Fermi liquids. Generically, NMR and NQR spin-lattice relaxation time $T_1$ is nearly $T$ and doping independent in the normal state $T > T_c$ [3], in contrast to the Korringa law $T_1^{-1} \propto T$ in normal metals. Also, in the same regime the low-$\omega$ dynamical susceptibility in doped systems appears to be consistent with $\chi''(\omega) \propto \omega/T$ [1].

Well understood so far are only undoped cuprates, which behave in all respects as isotropic quantum AFM, with the long range order at $T = 0$. For doped systems NMR and NQR data on spin dynamics have been interpreted within the phenomenological model of AFM correlated spins, where the $T$ dependence is attributed to the variation of AFM correlation length $\xi(T)$ [4]. At low hole doping dynamics has been mapped to related quantum-critical (QC) scaling regime of nonlinear sigma model where $\xi \propto 1/T$ [5]. An alternative scenario for the low-$\omega$, low-$T$ behavior has been given in terms of an anomalous T-dependence (not directly related to $\xi(T)$), introduced within the marginal Fermi-liquid hypothesis [6].

Among the microscopical models the most helpful for the discussion of spin dynamics in strongly correlated metals is the $t - J$ model [7]

$$H = -t \sum_{\langle ij \rangle s} (c_{js}^\dagger c_{is} + H.c.) + J \sum_{\langle ij \rangle} (\vec{S}_i \cdot \vec{S}_j - \frac{1}{4} n_i n_j), \qquad (1)$$

where $c_{is}^\dagger(c_{is})$ are projected fermionic operators, taking into account that the double occupancy of sites is not allowed. In spite of its simple form the model proved to be very difficult to analyse, both analytically [7] as well as numerically [8]. Most reliable results related to spin dynamics have been obtained by exact diagonalization studies [8–10] of small systems, and by means of the high-temperature series expansion [11,12]. These methods, however, have not been able so far to get results for low-$\omega$ and d.c. spin response in the most challenging parameter regime $J < t$ and at intermediate $0 < T < J$.

Recently present authors introduced a new numerical method, based on the Lanczos diagonalization of small systems combined with the random sampling over basis states [13], which allows the study of dynamical and d.c. response functions at $T > 0$. The method has been already applied to the evaluation of optical and d.c. conductivity in the $t - J$ model on a square lattice [14].

In this paper we present results, obtained by the same method, for the dynamical spin-structure factor (we furtheron choose units with $\hbar = k_B = 1$)

$$S(\vec{q}, \omega) = \operatorname{Re} \int_0^\infty dt \, e^{i\omega t} \langle S_{\vec{q}}^z(t) S_{-\vec{q}}^z(0) \rangle, \qquad (2)$$

and related dynamical susceptibility $\chi(\vec{q}, \omega)$

$$\chi''(\vec{q}, \omega) = (1 - e^{-\beta\omega}) S(\vec{q}, \omega). \qquad (3)$$

Since numerical requirements are the same as for the conductivity problem, we refer to the description of technical details to Ref. [13,14]. We study the planar system of $N = 4 \times 4 = 16$ sites with $J/t = 0.3$ and variable doping, i.e. with $N_h = 0 - 10$ holes. Typically we use up to $M = 120$ Lanczos steps and the sampling over $N_0 \sim 300$ states. The method has been tested with the full-diagonalization results for $N = 10$ [9]. In comparison with latter results our discrete spectra obtained for $N = 16$ are much more dense, so that even minimal additional broadening (with a characteristic width $\eta \ll J$) yields smooth macroscopic-like spectra. Generally, results become unrealistic at low temperatures $T < T^*$, where finite-size effects start to introduce size-dependent features. $T^*$ is related to the average level distance in the low-energy sector. The latter is smallest (for fixed $N$) in the intermediate-doping regime $0.12 < c_h < 0.5$, where we reach $T^* \sim 0.1t$. The low-energy sector becomes more sparse both in the low-electron $c_h > 0.5$ as well as in the pure AFM regime $c_h < 0.12$, leading to the increase of $T^*$. The onset of finite-size effects is monitored by the appearance of unphysical structures in the high-frequency spectra, by the dependence on smoothing $\eta$ etc. We choose furtheron predominantly $\eta = 0.07 \, t$.

We present in Fig. 1 $S(\vec{Q},\omega)$ spectra at the AFM wavector $\vec{Q} = (\pi,\pi)$ and fixed $T = 0.2\ t < J$, but varying hole concentration $c_h$. As noted above, moderate doping results are reliable in this $T < J$ regime, while low-doping spectra already exhibit finite-size effects, e.g. gaps appearing in spectra etc. The most interesting feature in Fig. 1 is the qualitative change of spectra on doping. While at $c_h < 0.12$ spectra are dominated by a single central peak with the width $\omega \sim 2J$ due to AFM fluctuations, in the intermediate regime $0.12 < c_h < 0.5$ *a high-frequency component with $\omega \sim t$ emerges, coexisting with the remaining low-frequency fluctuations*. It is plausible to attribute the high-$\omega$ dynamics to the free-fermion-like component of the correlated system, in particular since it appears to be quite independent on $J$ provided that $J < t$. Also the free-fermion part seems to exhaust the spectra in the overdoped cases $c_h > 0.5$. Although this observation is not unexpected [9,15], the coexistence of spin-fluctuation and free-fermion timescales at optimum doping has been established here for the first time. It is namely harder for other methods, e.g. the high-$T$ expansion method, to treat coexisting different timescales. The dual character is a crucial property, since the free-fermion part determines to large extent static spin correlations $S(\vec{q})$ (and charge density correlations $N(\vec{q})$), interpreted in terms of quasi-Fermi surface [15]. On the other hand the low-$\omega$ spin dynamics dominates dynamical and static spin susceptibilities, $\chi''(\vec{q},\omega)/\omega$ and $\chi(\vec{q})$, respectively, hence also the neutron scattering and NMR processes.

Figure 2 displays dynamical spectra $\chi''(\vec{q},\omega)/\omega$ for fixed $c_h = 3/16$, but various $T$ and nonequivalent $\vec{q}$ (note that on the $4 \times 4$ lattice $\vec{q} = (0,\pi)$ and $\vec{q} = (\pi/2,\pi/2)$ are equivalent). In contrast to Fig. 1, high-$\omega$ features are suppressed here. Nevertheless, the free-fermion part is well separated from the low-$\omega$ part only for $\vec{q} \sim \vec{Q}$, for which one expects a gap in the response of free fermions with a well defined Fermi surface. For other $\vec{q}$ (at given low doping) the free-fermion contribution persists at larger $\omega > J$ in the form of a long tail, while in the low-$\omega$ regime it merges with the spin contribution. The most striking feature of Fig. 2 is however *strong $T$ dependence of low-$\omega$ spectra, whereas at the same time the AFM correlation length $\xi$ is at most only weakly $T$ dependent*. Here $\xi(T)$ can be estimated from the static $\chi(\vec{q})$ or from the ratio of spectra with different $\vec{q}$. The above conclusion seems to hold for all $|\vec{q} - \vec{Q}| < \alpha \xi^{-1}$, where $\chi(\vec{q}) \sim \chi(\vec{Q})$. The relevant volume in the $\vec{q}$-space clearly increases on doping and exhausts for $c_h = 3/16, 4/16$ already the majority of the Brillouin zone, while within the same doping regime scaling does not hold for e.g. $\vec{q} = (0, \pi/2)$. The variation at $\omega < J$ is $\chi''(\vec{q}, \omega)/\omega \propto 1/T$, or *equivalently $S(\vec{q}, \omega)$ is nearly independent on $T$ and $\omega$ at $\omega < J$* in the same regime, as also observed in neutron-scattering experiments [1], where $\chi''(\omega) \propto \omega/T$. It should be noted, that the universal scaling $\chi''(\omega) = f(\omega/T)$ claimed by several authors [1] seems to be close to the requirement of $S(\vec{q}, \omega) \sim const$.

Spectra discussed above have as a direct consequence $T$-variation of static $\chi(\vec{q})$ also at $T < J$. We observe pronounced $T$ dependence, e.g. $\chi(\vec{q}) \propto T^{-1}$ for $c_h = 3/16$, in a wide regime $J/3 < T < t$ for all $q$ within the correlation volume. It should be however noted that we are quite restricted in the range of $T/J$, so that more quantitative conclusions on possible power-law (or logarithmic) variation with $T$ are not feasible.

We can discuss our results in relation to experimental ones obtained in cuprates via the NMR and NQR relaxation. The NQR spin-lattice relaxation time $T_1$ and the spin-echo decay time $T_2$ for $^{63}$Cu nuclei are related to electronic spin susceptibilities by [4,3]

$$\frac{1}{T_1} = \frac{2T}{g^2 \mu_B^2} \frac{1}{N} \sum_{\vec{q} \neq 0} A_\perp^2(\vec{q}) \frac{\chi''(\vec{q}, \omega_0)}{\omega_0}, \qquad \omega_0 \to 0,$$

$$\frac{1}{T_2} = \sqrt{\frac{0.69}{8}} \frac{1}{g^2 \mu_B^2} \left[ \frac{1}{N} \sum_{\vec{q}} A_\parallel^4(\vec{q}) \chi^2(\vec{q}) - \left( \frac{1}{N} \sum_{\vec{q}} A_\parallel^2(\vec{q}) \chi(\vec{q}) \right)^2 \right]^{1/2}. \qquad (4)$$

Note that in the evaluation of $T_1$ we neglect the $q \sim 0$ spin-diffusion contribution, which is supposed to be less important [10]. To allow a direct comparison with experiments we choose $A_\perp(\vec{q}), A_\parallel(\vec{q})$ as proposed in the literature [4] and $t = 0.4~eV$ [7]. Again, $J = 0.3~t = 0.12~eV$.

Results for $T_1$ and for the ratio $R = T_1 T/T_2$ are presented on Figs. 3a,b. For the undoped case our results for $T_1$ agree with Ref. [10], but the $T$-variation for $T < J$ is already influenced by finite-size effects (due to sparse density of states at low energies). It is remarkable, that $T_1$ appears to be nearly $T$ independent for a broad range of hole concentrations $0.06 < c_h \leq 0.31$ ( $c_h = 6/16$ is not included due to more pronounced finite-size effects caused by a closed-shell configuration) . Only for the overdoped systems with $c_h \geq 0.5$ the behavior for $T < t$ approaches that of a normal Fermi liquid with $T_1^{-1} \propto T$. Our results are in agreement, even quantitatively without any fitting parameters, with recent remarkable NQR experiments in $La_{2-x}Sr_xO_4$ [3], which reveal nearly $T$ and $x$ independent $T_1$ for $T > 300K$ and $x < 0.15$. We in fact establish a variation of $T_1$ with doping, which however becomes more pronounced only for $c_h > 2/16$. For optimum doping lower rates $T_1^{-1}$ are expected, consistent with the data for $YBa_2Cu_3O_7$ [2], where for $T > T_c$ $T_1$ is again found to be only weakly dependent on $T$.

Temperature independent ratio $R = T_1T/T_2$, as approximately realized in cuprates, has been interpreted as the evidence for the QC behavior of the effective spin system [5]. We find quite analogous weak $T$ variation of the ratio within the $t - J$ model, with results presented in Fig. 3b. Here undoped $c_h = 0$ case is omitted due to inaccurate (finite-size dominated) results for $\chi(\vec{Q})$ and consequently for $T_2$ obtained on a small system for $T < J$. The origin of $R(T) \sim const$ is however considerably different from the QC scenario, since the $T_2(T)$ dependence is not connected (in an evident way) with the $\xi(T)$ variation. Results indicate on a stronger doping dependence, even at low doping. Quantitatively obtained values are in reasonable agreement with experimental ones, e.g. $R \sim 1700K$ at $T = 300K$ for $YBa_2Cu_3O_7$ ($c_h \sim 0.23$), while $R \sim 2400K$ for $YBa_2Cu_3O_{6.63}$ [5].

In conclusion, we have presented results for the dynamical spin susceptibility obtained within the $t - J$ model with the new numerical method, for the first time in the challenging regime of low to moderate doping with $J < t$, as well as $T, \omega < J$. The most interesting finding is the anomalous low-frequency spin dynamics, showing up also in the $T$ dependence of static susceptibilities, which is related to NMR relaxation times $T_1$ and $T_2$, as well as to neutron scattering experiments. Our results are consistent with nearly $T$- and $\omega$-independent $S(\vec{q}, \omega)$ (or $\chi''(\vec{q}, \omega) \propto \omega/T$) for $\vec{q}$ around $\vec{Q}$, even in the regime where $\xi \neq \xi(T)$, but still well in the normal state. Such $T$ dependence is close to the concept of the marginal Fermi liquid [6].

It is tempting to speculate on the origin of the anomalous low-$\omega$ spin dynamics. It is quite plausible to relate it to the dramatic increase of the density of low-energy many-body states emerging on doping. The latter shows up in substantially enhanced entropy at low $T$ [16], and in the strong carrier scattering as manifested in the resistivity $\rho \propto T$, established recently also within the $t-J$ model [14]. An intuitive picture might be that at low $T < J$ spin clusters with the characterictic size $l \sim \xi(T)$ behave as nearly independent whereby their interaction is effectively blocked by doped holes, thus leading to large density of low-$\omega$ modes. A more coherent theoretical description is clearly missing so far. Our study shows that the $t-J$ model still remains a promising starting point for these investigations.

This work has been supported by the Ministry of Science and Technology of Slovenia. One of the authors (P.P.) wishes to thank R.R. Singh, H. Monien, T.M. Rice and D. Pines for fruitful discussions and suggestions, as well as to the Los Alamos National Laboratory for the support and hospitality during the completion of the work.

**Figure captions**

Fig. 1. Dynamical spin-structure factor $S(\vec{q}=\vec{Q},\omega)$ at various hole dopings and fixed $T = 0.2\ t < J$. Spectra for planar 16-site system are broadened with $\eta = 0.07\ t$.

Fig 2. Dynamical spin susceptibility $\chi''(\vec{q},\omega)/\omega$ at fixed doping $c_h = 3/16$ for different $\vec{q}$ in the Brillouin zone and various temperatures: $T/t = 0.1$ (full line), 0.2 (dashed line), 0.3 (dash-dotted line) and 0.5 (dotted line).

Fig 3. a) NQR spin-lattice relaxation rate $1/T_1$ and b) the ratio $T_1T/T_2$, both vs. $T/t$ for various dopings $c_h$. Note that here $t = 0.4\ eV = 4640\ K$.